\def\preprint{TUW--27--99}       
\def\finished{December 1999}
\def\archive {hep-th/9912174}           

\def\Title{	$\Lambda$-symmetry and background independence of \\[4mm]
	noncommutative gauge theory on $\mathbb R^n$}

\long\def\Abstract{
\hspace*{18pt}%
Background independence of noncommutative Yang-Mills theory on $\mathbb R^n$ 
is discussed. The quantity $\theta \hat F \theta - \theta$ is found to be 
background dependent at subleading order, and it becomes background 
independent only when the ordinary gauge field strength $F$ is constant. It
is shown that, at small values of $B$, the noncommutative Dirac-Born-Infeld
action possesses $\Lambda$-symmetry at least to subleading order in $\theta$
if $F$ damps fast enough at infinity.
}

\documentclass[12pt]{article}
\usepackage[latin1]{inputenc}
\usepackage{amsmath}
\makeatletter

\providecommand{\LyX}{L\kern-.1667em\lower.25em\hbox{Y}\kern-.125emX\@}

\usepackage{latexsym}
\makeatother

\usepackage{epic,epsf,amssymb,latexsym}

\def\ifundefined#1{\expandafter\ifx\csname#1\endcsname\relax}
\def\bye{\end{document}}   \def\HS#1 {\hspace*{#1pt}}

\def\BC{\begin{center}}    
\def\EC{\end{center}}       

\def\fnote#1#2{\begingroup\def\thefootnote{#1}\footnote{#2}
                \addtocounter{footnote}{-1}\endgroup}   

\let\0=\over      \let\1=\vec      \def\2{{1\over2}}    \let\nn=\nonumber
\let\4=\underline \let\5=\overline \let\6=\partial      \def\7#1{{#1}\llap{/}}
\def\8#1{{\textstyle{#1}}}         \def\9#1{{\ifmmode{\pmb{#1}}\else\bf#1\fi}}

          \let\d=\delta   
         \let\th=\theta

\def\mao#1{\mathop{\rm #1}\nolimits}    \def\Tr{\mao{Tr}}

\begin{document}

\vspace*{-58pt}\begin{flushright}  \archive\\[7pt] \preprint  \end{flushright}
\vspace{-3mm}

\BC{\LARGE\bf \Title}
\\[17mm]{\large
        Maximilian KREUZER\fnote{*}{e-mail: kreuzer@hep.itp.tuwien.ac.at}
        and Jian-Ge ZHOU\fnote{\#}{e-mail: jgzhou@hep.itp.tuwien.ac.at}	}
\\[5mm]
        Institut f\"ur Theoretische Physik, Technische Universit\"at Wien\\
        Wiedner Hauptstra\ss e 8--10, A-1040 Wien, AUSTRIA
        
\vfill          {\bf ABSTRACT } \\[9mm]	\parbox{150mm}{ \Abstract}	\EC

\vfill \noindent \preprint \\[5pt] \finished \vspace*{9mm} \baselineskip=21pt

\setcounter{page}{0} \thispagestyle{empty} \newpage \pagestyle{plain}  

\section{Introduction}

Recently, Seiberg and Witten observed that ordinary and noncommutative gauge
fields can be induced by the same $2D$ $\sigma$-model action regularized
in different ways \cite{sw1}. Then they argued that ordinary gauge fields
should be equivalent to noncommutative gauge theory, which, in a certain limit,
acts as the low energy effective theory of open stings. Furthermore,
they conjectured that the noncommutative effective Lagrangian takes the 
same form as the ordinary Dirac-Born-Infeld (DBI)
action for the \( D \)-brane, except that the product of functions is replaced
by the star product and Lorentz indices are contracted by the open-string 
metric $G$ instead of the closed-string metric $g$. The explicit transformation
between the ordinary gauge field \( A \) and the noncommutative one \(\hat{A}\)
was given up to order \( \theta  \) \cite{sw1}. 
Even at this order there is an ambiguity in the general transformation 
between $A$ and $\hat A$, 
which, however, 
can be removed by a field redefinition of the noncommutative gauge field 
\cite{ak2}.

The classical \( \sigma  \)-model action in a
$B$-field background has two 
abelian symmetries: One is the $U(1)$ gauge 
symmetry \( A\rightarrow A+d\, \lambda  \), 
and the other is the ``\( \Lambda  \)-symmetry''
\( A\rightarrow A-\Lambda ,\, B\rightarrow B+d\Lambda  \),
which keeps \( B+F \) invariant. In \cite{w3} it was argued that 
the $U(1)$ gauge 
symmetry
is enhanced to a \( U( N)  \)  when \( N\, \, \, D \)-branes
coincide, while the \( \Lambda  \)-symmetry 
is believed to 
act on the \( U\left( 1\right)  \) part of the \( U\left( N\right)  \)
symmetry. 
In the ordinary DBI Lagrangian, the \( \Lambda  \)-symmetry is manifest.
So it is natural to ask, whether the noncommutative gauge theory has 
\( \Lambda  \)-symmetry
or not if its ordinary counterpart has this symmetry.

In ref. \cite{sw1} background independence of noncommutative Yang-Mills 
theory
on \( \mathbb R^{n } \) was considered. In ordinary Yang-Mills theory, the 
gauge-invariant
combination of \( B \) and \( F \) is \( M=B+F \). The same gauge invariant
field \( M \) can be split in different ways as \( B+F \) or 
\( B' +F'  \).
In \cite{sw1}, it was shown that the noncommutative Yang-Mills theory on 
\( \mathbb R^{n} \)
is invariant under those different splitting, which was called background 
independence
of the noncommutative Yang-Mills theory. 
Of course this 
is nothing but the \( \Lambda  \)-symmetry in the context of the noncommutative
Yang-Mills theory. What Seiberg and Witten did in \cite{sw1} is that they 
analysed
the \( \Lambda  \)-symmetry in the framework of ordinary Yang-Mills theory
on noncommutative spacetime. Since the ordinary gauge theory on noncommutative
spacetime is supposed to be equivalent to noncommutative
gauge theory on commutative (i.e., ordinary) spacetime \cite{cds4}, it
is interesting to see how \( \Lambda  \)-symmetry is realized in 
noncommutative Yang-Mills theory on ordinary spacetime.

In the present paper we discuss the background independence of 
noncommutative
Yang-Mills theory on \( \mathbb R^{n} \) from the point of view of the 
noncommutative
gauge theory on ordinary spacetime. We first give the explicit expression for
the transformation between ordinary \( U\left( 1\right)  \) gauge fields and
noncommutative ones up to second order in $\theta$,%
\footnote{
	~The expression for \( \hat{A} \) in terms of \( A \) up to order 
	\( \theta ^{2} \)  was given in \cite{o5}.} 
which can be obtained from the differential equation given in
\cite{sw1}. We check the
\( \Lambda  \)-symmetry of the noncommutative Yang-Mills theory on 
\( \mathbb R^{n} \)
up to subleading order. We find that for the noncommutative Yang-Mills 
field
on ordinary spacetime, the quantity \( Q=\theta \hat{F}\theta -\theta  \) is
background dependent for general \( F \), and only when \( F \) is constant
we have \( Q=-\left( B+F\right) ^{-1} \) which is background independent.
The measure \( d^{n}x \sqrt{G}/g_{YM}^{2} \) varies with \( \theta , \)
but the noncommutative Yang-Mills action is invariant under 
\( \Lambda  \)-symmetry
if the ordinary field strength \( F \) damps fast enough at infinity. We 
further
show that the noncommutative DBI action has \( \Lambda  \)-symmetry in the
case of small \( B \) (i.e., small \( \theta  \)) up to subleading order.
All of this indicates that noncommutative gauge theories,
which act as low energy effective theories  of open strings, 
indeed possess \( \Lambda  \)-symmetry.

The paper is organized as follows. In the next section, we first give the 
expression
for the noncommutative field strength \( \hat{F}{( \hat{A}) } \) in
terms of the ordinary \( A \) and \( F \) up to order \( \theta ^{2} \).
We examine the quantity \( Q \) and the measure \( d^{n}x\sqrt{G}/g_{YM}^{2}\)
in the context of noncommutative gauge theory on ordinary spacetime. We show
that the noncommutative Yang-Mills action is background independent. In section
3, we check \( \Lambda  \)-symmetry for the noncommutative DBI action 
for small values of \( B. \) In section 4, we present our summary and 
discussion.

\section{Background independence of noncommutative Yang-Mills theory on 
	$\mathbb R^{n}$}

Demanding that ordinary gauge fields that are gauge-equivalent are mapped
to noncommutative gauge fields that are likewise gauge-equivalent, the 
transformation
between the noncommutative gauge field \( \hat{A} \) and the ordinary \( A \)
can be described by \cite{sw1}
\begin{equation}
\label{1}
\hat{A}\left( A\right) +\delta _{\hat{\lambda }}\, 
	\hat{A}\left( A\right) =\hat{A}\left( A+\delta _{\lambda }A\right) 
\end{equation}
with infinitesimal \( \lambda  \) and \( \hat{\lambda }\, 
	\left( \lambda ,A\right) . \)

The equation (\ref{1}) can be solved by expanding it in powers of the 
noncommutative
parameter \( \theta  \). The differential equations, which describe 
	how \( \hat{A}\left( \th\right)  \)
and \( \hat{\lambda } \)\( \left( \th\right)  \) should change when 
\( \theta  \) varies, are given by \cite{sw1}
\begin{eqnarray}
\delta \hat{A}_i\left( \theta \right)  & = & -\frac{1}{4}\delta 
\theta ^{kl}\left[ \hat{A}_{k}*\left( \partial _{l}\hat{A}_{i}
	+\hat F_{li}\right) 
+\left( \partial _{l}\hat{A}_{i}+\hat{F}_{li}\right) *\hat{A}_{k}\right] 
\label{2} \\
\delta \hat\lambda \left( \theta \right) & = & \frac{1}{4}\delta \theta ^{kl}\,
\left( \partial _{k}\hat{\lambda }*\hat{A}_{l}+\hat{A}_{l}*\partial _{k}\, 
\hat{\lambda }\right) \label{3} \\
\delta \hat{F}_{ij}(\theta)  & = & \frac{1}{4}\delta \theta ^{kl}
	\left[ 2\hat{F}_{ik}*\hat{F}_{jl}+2\hat{F}_{jl}*
	\hat{F}_{ik}-\hat{A}_{k}*\left( \hat{D}_{l}\, \hat{F}_{ij}
	+\partial _{l}\hat{F}_{ij}\right) \right. \nonumber \\
 &  & \left. -\left( \hat{D}_{l}\, \hat{F}_{ij}
	+\partial _{l}\hat{F}_{ij}\right) *\hat{A}_{k}\right] \label{4} 
\end{eqnarray}
Replacing \( \hat{A} \) by \( A \) on the right side of the above equations,
we get the solutions up to order \( \theta  \), and for 
\( U\left( 1\right)  \) gauge fields we have.%
\footnote{
	~We are interested in \( \hat{A} \), especially in \( \hat{F} \), 
	so we do not calculate \( \hat{\lambda } \).}
\begin{eqnarray}
\hat{A}_{i}^{\left( 1\right) } & = & A_{i}-\frac{1}{2}\theta ^{kl}A_{k}
	\left( \partial _{l}A_{i}+F_{li}\right) +\mathcal{O}\left( \theta ^{2}
	\right) \nonumber \\
\hat{F}_{ij}^{\left( 1\right) } & = & F_{ij}
	-\theta ^{kl}\left( F_{ik}\, F_{lj}+A_{k}\partial _{l}F_{ij}\right) 
	+\mathcal{O}\left( \theta ^{2}\right) \label{5} 
\end{eqnarray}
Inserting \( \hat{A}^{(1)} \) and \( \hat{F}^{(1)} \) into the
right hand side of equations (\ref{2}) and (\ref{4}) we obtain \( \hat{A} \) 
and \( \hat{F} \) up to order \( \theta ^{2} \), 
\begin{eqnarray}
\hat{A}_{i} & = & A_{i}-\frac{1}{2}\theta ^{kl}A_{k}\left( \partial _{l}A_{i}
+F_{li}\right) +\nonumber \\
 &  & \frac{1}{2}\theta ^{kl}\theta ^{mn}\left\{ A_{k}
	\left[ \partial _{l}A_{m}	\partial _{n}A_{i}
	-\left( \partial _{l}F_{mi}\right) A_{n}+F_{lm}F_{ni}\right]\right\}
	+\mathcal{O}\left( \theta ^{3}\right), \label{6} \\[3mm]
\label{7}
\hat{F}&=&F-F\theta F-\theta ^{kl}A_{k}
\partial _{l}F+\mathcal{T}_{\theta ^{2}}+\mathcal{O}\left( \theta ^{3}\right) 
\end{eqnarray}
with
\begin{equation}
\label{8}
\mathcal{T}_{\theta ^{2}}=F\theta F\theta F+\frac{1}{2}\,A_{k}\theta ^{kl}
	\left( \partial _{l}A_{m}+F_{lm}\right) \theta ^{mn}\partial _{n}F
	+\th^{kl}A_k\6_l(F\th F)+\2\,\th^{kl}\th^{mn}A_kA_m\6_l\6_n F.
\end{equation}
To obtain eqs. (\ref{6}-\ref{8}), we solved eqs. (\ref{2}) and (\ref{4}) 
perturbatively, and we 
treat the ordinary \( F \) as \( \theta  \) independent, which is
different from the following discussion about \( \Lambda  \)-transformations,
where \( F \) will vary with \( \theta  \).

Now let us consider  noncommutative Yang-Mills theory on 
\( \mathbb R^{n } \). The action is \cite{sw1}
\begin{equation}
\label{9}
\hat{L}=\frac{1}{g_{YM}^{2}}\int d^{n}x\sqrt{G}G^{ik}G^{jl}\left( \hat{F}_{ij}
	-\theta _{ij}^{-1}\right) *\left( \hat{F}_{kl}-\theta _{kl}^{-1}\right)
\end{equation}
with \( G \) and \( \theta  \) defined by
\begin{equation}
\label{10}
\theta =B^{-1}, ~~~~~~~~~ G=-Bg^{-1}B,
\end{equation}
where we have put \( 2\pi \alpha'  =1 \) for simplicity.
Ignoring total derivatives and constant terms  we rewrite
the action (\ref{9}) as
\begin{equation}
\label{11}
\hat{L}=-\frac{1}{g_{YM}^{2}}\int d^{n}x
	\sqrt{G}\Tr\left(\hat{Q}g\hat{Q}g\right)
\end{equation}
with
\begin{equation}
\label{12}
\hat{Q}=\theta \hat{F}\theta 
\end{equation}
Hence it is sufficient to show that the action (\ref{11}) is background 
independent. 

Consider the \( \Lambda  \)-transformation, defined by
\begin{equation}
\label{13}
F\rightarrow F-\delta B=F+\theta ^{-1}\delta \theta \theta ^{-1},
\end{equation}
which keeps \( F+B \) invariant. From (\ref{10}) we know that under a 
\( \Lambda  \)-transformation
\( \theta  \) changes as \( \theta \rightarrow \theta + \delta \theta  \), 
while the ordinary field strength \( F \) transforms as \( F\rightarrow F
	+\theta ^{-1}\delta \theta \theta ^{-1}. \)
Here we should point out that, even for small $\theta$, we will assume 
\( \delta \theta \ll \theta  \),
and of course we only need to keep terms linear in \( \delta \theta  \).

Under a \( \Lambda  \)-transformation  \( \hat{F} \) changes as
\begin{eqnarray}
\hat{F}\left( \theta +\delta \theta \right)  & = & F+\theta ^{-1}\delta \theta
 \theta ^{-1}-\left( F+\theta ^{-1}\delta \theta \theta ^{-1}\right) 
	\left( \theta +\delta \theta \right) \left( F+\theta ^{-1}\delta 
	\theta \theta ^{-1}\right) \nonumber \\
 & - & \left[ A_{k}-\frac{1}{2}\left( \theta ^{-1}\delta \theta 
	\theta ^{-1}\right) _{ks}\, x ^{s}\right]
	 \left( \theta +\delta \theta \right) ^{kl}\partial _{l}F\nonumber \\
 & + & \mathcal{T}_{\left( \theta +\delta \theta \right) ^{2}}
	+\mathcal{O}\left[ \left( \theta +\delta \theta \right) ^{3}\right] 
	\label{14} 
\end{eqnarray}
Collecting the terms linear in $\d\theta$ we thus obtain
\begin{eqnarray}
\delta \hat{F}\left( \theta \right)  & = & 
	\theta ^{-1}\delta \theta \theta ^{-1}-\theta ^{-1}\delta \theta F
	-F\delta \theta \theta ^{-1}\nonumber -F\delta \theta F\\
 & - & \frac{1}{2}\left( \delta \theta \theta ^{-1}\right)^l{} _{s}\, 
	x ^{s}\partial _{l}F-A_{k}\delta 
	\theta ^{kl}\, \partial _{l}F+H_{\delta \theta }+
	\hbox{higher order terms},
		\label{15} 
\end{eqnarray}
where \( H_{\delta \theta } \) represents the 
	\( \delta \theta  \)
order terms coming from \( \mathcal{T}_{\left( \theta +\delta \theta \right) 
	^{2}}-\mathcal{T}_{\theta ^{2}} \),
which is obtained by replacing one of the \( F \)'s by \( \theta ^{-1}\delta 
	\theta \theta ^{-1} \)
or an \( A_{i} \) by \( -\frac{1}{2}\left( \theta ^{-1}\delta \theta 
	\theta ^{-1}\right) _{ij}\,x^{j} \)
in \( \mathcal{T}_{\theta ^{2}} \). It has the form
\begin{eqnarray}
H_{\delta \theta } & = & \theta ^{-1}\delta \theta F\theta F+F\delta \theta F
	+F\theta F\delta \theta \, \theta ^{-1}\nonumber \\
 & + & \frac{1}{4}\left( \delta \theta \, \theta ^{-1}\right) ^l{}_k\, 
	x ^{k}\left( \partial _{l}A_{m}+F_{lm}\right) 
	\theta ^{mn}\partial _{n}F\nonumber \\
 & + & \frac{3}{4}A_{k}\delta \theta ^{kl}\partial _{l}F
	+\frac{1}{2}\left( \delta \theta \theta ^{-1}\right) ^l{}_k
	\,x^{k}\partial _{l}\left( F\theta F\right) \nonumber \\
 & + & A_{k}\theta ^{kl}\partial _{l}\left( \theta ^{-1}\delta \theta F
	+F\delta \theta \theta ^{-1}\right) +\frac{1}{2}
	\left( \delta \theta \theta ^{-1}\right) ^l{}_k\,x^{k}
	\theta ^{mn}A_{m}\partial _{l}\partial _{n}F\label{16} 
\end{eqnarray}
Inserting (\ref{16}) into (\ref{15}), then \( \delta \hat{F} \) can be 
written as
\begin{eqnarray}
\delta \hat{F}\left( \theta \right)  & = & \theta ^{-1}\delta \theta 
	\theta ^{-1}-\theta ^{-1}\delta \theta F-F\delta \theta \theta ^{-1}
	-\frac{1}{2}\left( \delta \theta \theta ^{-1}\right) ^l{}_k
	\,x^{k}\partial _{l}F\nonumber \\
 & - & \frac{1}{4}A_{k}\delta \theta ^{kl}\partial _{l}F
	+\theta ^{-1}\delta \theta F\theta F
	+F\theta F\delta \theta \theta ^{-1}\nonumber \\
 & + & \frac{1}{4}\left( \delta \theta \theta ^{-1}\right) ^l{}_k
	\,x^{k}\left( \partial _{l}A_{m}+F_{lm}\right) 
	\theta ^{mn}\partial _{n}F\nonumber \\
 & + & \frac{1}{2}\left( \delta \theta \theta ^{-1}\right) ^l{}_k
	\,x^{k}\partial _{l}\left( F\theta F\right) \nonumber \\
 & + & A_{k}\theta ^{kl}\partial _{l}\left( \theta ^{-1}\delta \theta F
	+F\delta \theta \theta ^{-1}\right) \nonumber \\
 & + & \frac{1}{2}\left( \delta \theta \theta ^{-1}\right) ^l{}_k
	\,x^{k}\theta ^{mn}A_{m}\partial _{l}\partial _{n}F, \label{17} 
\end{eqnarray}
where the leading order term is \( \theta ^{-1}\delta \theta 
	\theta ^{-1}\left( \theta ^{-2}\delta \theta \right) , \)
the subleading order term is \( \theta ^{-1}\delta \theta  \) and the third
order term is \( \delta \theta  \). Since the leading order term 
	\( \theta ^{-2}\delta \theta  \)
gives a trivial contribution to the variation of the noncommutative Yang-Mills
Lagrangian,  the subleading order term \( \theta ^{-1}\delta \theta  \)
is enhanced to the nontrivial leading order  and the \( \delta \theta  \) 
terms are lifted to subleading order.

Before proceeding, we would like to 
compare 
	\( \delta \hat{F}\left( \theta \right)  \)
defined in (\ref{17}) with 
the variation \( \delta \hat{F}\left( \theta \right)\)
in (\ref{4}), which describes the variation of \( \hat{F} \) under a 
change of the
noncommutative parameter \(\theta\) with the ordinary field strength \( F \)
kept fixed. 
Under a \( \Lambda  \)-transformation, on the other hand, 
\( \theta  \) changes, and the ordinary field strength varies simultaneously
to keep \( B+F \) invariant. In (\ref{17}), when we vary \( \theta  \) (induced
by changing \( B \)) we keep $g_{s}$ and $g$ fixed, and we are sticking with
point-splitting regularization. In \cite{sw1}, Seiberg and Witten argued that
one can use a suitable regularization which somehow interpolates between 
Pauli-Villars and point-splitting, then one can vary \( \theta  \) while 
holding \( g_{s},\, g \)
and \( B \) fixed. Thus  \( \delta \hat{F} \) defined 
in (\ref{17}) is conceptually different
from  (\ref{4}).

The variation of the quantity \( \hat{Q} \) defined in (\ref{12}) is 
\begin{eqnarray}
\delta \hat{Q} & = & \delta \theta -\frac{1}{2}\left( \delta \theta 
	\theta ^{-1}\right) ^l{}_k\,x^{k}\partial _{l}
	\left( \theta F\theta \right) +\left\{ -\frac{1}{4}A_{k}
	\delta \theta ^{kl}\partial _{l}\left( \theta F\theta \right) \right. 
\nonumber \\
 & + & \frac{1}{4}\left( \delta \theta \theta ^{-1}\right) ^l{}_k
	\,x^{k}\left( \partial _{l}A_{m}+F_{lm}\right) 
	\theta ^{mn}\partial _{n}\left( \theta F\theta \right) \nonumber \\
 & + & \left. \frac{1}{2}\left( \delta \theta \theta ^{-1}\right) ^l{}_k
	\,x^{k}\partial _{l}\left( \theta F\theta F\theta \right) 
	+\frac{1}{2}\left( \delta \theta \theta ^{-1}\right) ^l{}_k
	\,x^{k}A_{m}\theta ^{mn}\partial _{l}\partial _{n}
	\left( \theta F\theta \right) \right\}, \label{18} 
\end{eqnarray}
where the first term is of order \( \delta \theta  \), the second 
is of order \( \theta \delta \theta  \),
and the terms in the bracket are of the order of 
\( \theta ^{2}\delta \theta  \).

The noncommutative Yang-Mills action (\ref{11}) can be recast into
\begin{equation}
\label{19}
\hat{L}=\frac{1}{g_{s}}\int d^{n}x \, \det{}^{-\2}\theta \, 
	\Tr\left( \hat{Q}g\hat{Q}g\right),
\end{equation}
where we have exploited the \( \theta  \) dependence of 
	\( g_{YM}^{2} \) \cite{sw1} and
omitted an irrelevant numerical factor. Since we study the noncommutative 
Yang-Mills
theory on ordinary spacetime \(\mathbb R^{n}, \) the measure \( d^{n}x \) is invariant
under the change of \( \theta  \). Then the variation of the noncommutative
Yang-Mills action is
\begin{equation}
\label{20}
\delta \hat{L}=\frac{2}{g_{s}}\int d^{n}x ~\det{}^{-\2}\theta ~\delta S
	~+~\hbox{higher order terms}
\end{equation}
with
\begin{equation}
\label{21}
\delta S=\delta S_{1}+\delta S_{2}=-\frac{1}{4}\left( \Tr\theta ^{-1}\delta 
	\theta \right) \Tr\left( \hat{Q}g\hat{Q}g\right) 
	+\Tr\left[ \left( \delta \hat{Q}\right) g\hat{Q}g\right] 
\end{equation}
where \( \delta S_{1} \) and \( \delta S_{2} \) represent the leading and the
subleading order, respectively.

Since \( \hat{F} \) is a total derivative for \( U(1) \)
gauge fields, \( \delta \theta  \)
in \( \delta Q \) can be thrown away up to a total derivative. 
The leading order
term in \( \delta S \) is of order  \( \theta ^{3}\delta \theta , \) and
the subleading order  is \( \theta ^{4}\delta \theta  \).

At first, let us examine the leading order term \( \delta S_{1} \) with
\begin{eqnarray}
\delta S_{1} & = & -\frac{1}{4}\Tr\theta ^{-1}\delta \theta \, 
	\Tr\left( \theta F\theta g\theta F\theta g\right) \nonumber \\
 &  & -\frac{1}{2}\Tr\left\{ \left( \delta \theta \theta ^{-1}\right) ^l{}_k
	\,x^{k}\left[ \partial _{l}\left( \theta F\theta \right) \right] 
	g\theta F\theta g\right\} \nonumber \\
 & = & -\frac{1}{4}\Tr\, (\theta ^{-1}\delta \theta)\,\Tr\left(\theta F\theta 
	g\theta F\theta g\right) \nonumber \\
 &  & -\frac{1}{4}\Tr\left\{ \left( \delta \theta \theta ^{-1}\right) ^l{}_k
	\,x^{k}\left[ \partial _{l}\left( \theta F\theta g\right) ^{2}
	\right] \right\} \nonumber \\
 & = & \hbox{total derivative}\label{22} 
\end{eqnarray}
where from the second step to the last one we have integrated by parts. 
Eq. (\ref{22}) indicates that the noncommutative Yang-Mills action is 
background independent at leading order if the ordinary 
gauge field \( A \) damps fast enough at the infinity.

Next we consider the subleading term \( \delta S_{2} \), which is of order
\( \theta ^{4}\delta \theta  \),
\begin{eqnarray}
\delta S_{2} & = & -\frac{1}{4}\Tr\theta ^{-1}\delta \theta \times 2\Tr
	\left[ \theta F\theta g\theta \left( -F\theta F-A_{k}\theta ^{kl}
	\partial _{l}F\right) \theta g\right] \nonumber \\
 &  & +\Tr\left[ -\frac{1}{2}\left( \delta \theta \theta ^{-1}\right) ^l{}_k
	\,x^{k}\partial _{l}\left( \theta F\theta \right) g\theta 
	\left( -F\theta F-A_{m}\theta ^{mn}\partial _{n}F\right) 
	\theta g\right] \nonumber \\
 & + & \Tr\left\{ \left[ -\frac{1}{4}A_{k}\delta \theta ^{kl}
	\partial _{l}\left( \theta F\theta \right) +\frac{1}{4}
	\left( \delta \theta \theta ^{-1}\right) ^l{}_k\,x^{k}
	\left( \partial _{l}A_{m}+F_{lm}\right) \theta ^{mn}\partial _{n}
	\left( \theta F\theta \right) \right. \right. \nonumber \\
 & + & \left. \left. \frac{1}{2}\left( \delta \theta \theta ^{-1}
	\right) ^l{}_k\,x^{k}\partial _{l}\left( \theta F\theta F
	\theta \right) +\frac{1}{2}\left( \delta \theta \theta ^{-1}
	\right) ^l{}_{k}\,x^{k}A_{m}\theta ^{mn}\partial _{l}
	\partial _{n}\left( \theta F\theta \right) \right]
	g\theta F\theta g \right\}. \label{23}
\end{eqnarray}
Putting similar terms together we have
\begin{eqnarray}
\delta S_{2} & = & \frac{1}{2}\left( \delta \theta 
	\theta ^{-1}\right) ^l{}_k\,x^{k}A_{m}\theta ^{mn}\partial _{l}
	\Tr\left[ \partial _{n}\left( \theta F\theta g\right) 
	\theta F\theta g\right] \nonumber  \\
 & - & \frac{1}{8}\left( \Tr\theta ^{-1}\delta \theta \right) 
	\left( \Tr\theta F\right) \Tr\left( \theta F\theta g\theta F\theta 
	g\right) \nonumber \\
 & + & \frac{1}{16}\left( \Tr\delta \theta F\right) \Tr\left( \theta F\theta 
	g\theta F\theta g\right) \nonumber \\
 & + & \frac{1}{4}\left( \delta \theta \theta ^{-1}\right) _{\, k}^{l}
	\,x^{k}\left( \partial _{m}A_{l}\right) \theta ^{mn}
	\partial _{n}\left[ \Tr\left( \theta F\theta g\theta F\theta g\right) 
	\right] \nonumber \\
 & - & \frac{1}{8}\left( \delta \theta \theta ^{-1}\right) _{\, k}^{l}
	\,x^{k}\left( \partial _{m}A_{l}\right) \theta ^{mn}
	\partial _{n}\left[ \Tr\left( \theta F\theta g\theta F\theta g\right) 
	\right] \label{24} 
\end{eqnarray}
After integration by parts, we get 
\begin{equation}
\label{25}
\delta S_{2}=\hbox{total derivative}
\end{equation}
 which shows that under \( \Lambda  \)-transformation the noncommutative 
Yang-Mills
Lagrangian is invariant up to a total derivative also at the subleading order.

To compare our result with those in ref. \cite{sw1}, we define the quantity 
\( Q \)
by
\begin{equation}
\label{26}
Q=\hat{Q}-\theta =\theta \hat{F}\theta -\theta 
\end{equation}
Under a \( \Lambda  \)-transformation, the variation of \( Q \) can be obtained
from (\ref{18}) and is given by
\begin{eqnarray}
\delta Q & = & -\frac{1}{2}\left( \delta \theta \theta ^{-1}\right) ^l{}_k
  \,x^{k}\partial _{l}\left( \theta F\theta \right) -\frac{1}{4}A_{k}
  \delta \theta ^{kl}\partial _{l}\left( \theta F\theta \right) \nonumber \\
 & + & \frac{1}{4}\left( \delta \theta \theta ^{-1}\right) ^l{}_k
	\,x^{k}\left( \partial _{l}A_{m}+F_{lm}\right) 
	\theta ^{mn}\partial _{n}\left( \theta F\theta \right) \nonumber \\
 & + & \frac{1}{2}\left( \delta \theta \theta ^{-1}\right) ^l{}_k
	\,x^{k}\partial _{l}\left( \theta F\theta F\theta \right) 
	+\frac{1}{2}\left( \delta \theta \theta ^{-1}\right)^l{}_k
	\,x^{k}A_{m}\theta ^{mn}\partial _{l}\partial _{n}
	\left( \theta F\theta \right) \nonumber \\
 & + & \hbox{higher order terms}\label{27} 
\end{eqnarray}
which shows that the quantity \( Q \) is not invariant under 
\( \Lambda  \)-transformation
for general ordinary field \( A \) up to the subleading order term, but when
\( F \) is constant, we see \( \delta Q=0+\hbox{higher order terms} \), and in
this case \( Q \) can be expressed as \cite{sw1}
\begin{equation}
\label{28}
Q=-\frac{1}{B+F},
\end{equation}
where \( F \) is constant.

What we have learned from the above discussion is that when we 
consider the noncommutative
Yang-Mills theory on ordinary spacetime, the quantity \( Q \) generally is
not background independent, and the measure
\begin{equation}
\label{29}
\frac{d^{n}x \sqrt{G}}{g_{YM}^{2}}
\end{equation}
should also be background dependent. However, the total noncommutative 
Yang-Mills
action is background independent if the ordinary gauge field damps fast enough
at infinity. Because of
 the equivalence between ordinary gauge theory on 
noncommutative spacetime and noncommutative gauge theory on commutative 
(i.e., ordinary) spacetime \cite{cds4}, we believe 
that in the context of ordinary gauge theory
on noncommutative spacetime, the quantity \( Q \) and the measure 
\( d^{n}x \sqrt{G}/g_{YM}^{2} \)
are background independent individually, 
which is what Seiberg and Witten showed
in ref. \cite{sw1}.

\section{\protect\( \Lambda \protect \)-symmetry in the noncommutative DBI 
theory}

In section 2 we have analyzed the \( \Lambda  \)-symmetry of noncommutative
Yang-Mills theory, now we turn to the noncommutative DBI theory. The open 
string
metric and the noncommutative parameter \( \theta  \) are related to the 
closed-string
metric \( g \) and the constant background field \( B \) by \cite{sw1}
\begin{equation}
\label{30}
G=g-Bg^{-1}B,\,~~~~~~~ \, \theta =-\frac{1}{g+B}B\frac{1}{g-B}
\end{equation}
The explicit expression for the mapping between noncommutative gauge fields and
ordinary ones is valid only for small \( \theta  \), so we have
to restrict our discussion to this situation. From (\ref{30}) we see that there
are two possibilities to get a small value for \( \theta  \): 
Either  \( B \) is very large compared to \( g \), or it is much smaller than
\(g \). In the case of large \( B \), as argued in \cite{sw1}, the double
scaling limit should be imposed to make the model have proper sense, then the
corresponding noncommutative DBI theory is reduced to the noncommutative
Yang-Mills theory. Hence we only consider the noncommutative DBI theory
for small \( B \). Since at small \( B \)
\begin{equation}
\label{31}
\theta =-g^{-1}Bg^{-1}+\left( g^{-1}B\right) ^{3}g^{-1}+\mathcal{O}
	\left( B^{4}\right) 
\end{equation}
and the \( B^{2} \) term is absent, we will see below that 
it is good enough to use \( \theta =-g^{-1}Bg^{-1} \)
if we calculate to subleading order.
Then we have
\begin{equation}
\label{32}
	B=-g\theta g,~~~~~~~ G=g-g\theta g\theta g,
	~~~~~~~ G_{s}=g_{s}\det{}^{\frac{1}{2}}
	\left( 1 +g^{-1}B\right) 
\end{equation}
and their variations under a \( \Lambda  \)-transformation are
\begin{eqnarray}
\delta B & = & -g\delta \theta g,\, ~~~~~
	\delta G=-\left( g\delta \theta g\theta g
	+g\theta g\delta \theta g\right) \nonumber \\
\delta G_{s} & = & -\frac{1}{2}g_{s}\, \det{}^{\frac{1}{2}}\left( 1 
	+g^{-1}B\right) \, \Tr\left( \theta g\delta \theta g\right) \label{33} 
\end{eqnarray}
Under a \( \Lambda  \)-transformation \( F\rightarrow F+g\delta \theta g \)
and the noncommutative gauge field strength 
is
\begin{eqnarray}
\hat{F}\left( \theta +\delta \theta \right)  & = & F+g\delta \theta g
	-\left( F+g\delta \theta g\right) \left( \theta 
	+\delta \theta \right) \left( F+g\delta \theta g\right) \nonumber \\
 & - & \left[ A_{k}-\frac{1}{2}\left( g\delta \theta g\right) 
	_{ks}\,x^{s}\right] \left( \theta +\delta \theta \right) ^{kl}
	\partial _{l}F+\mathcal{T}_{\left( \theta +\delta \theta \right) ^{2}}
	\nonumber \\
 & + & \hbox{higher order terms}\label{34} \nonumber
\end{eqnarray}
The variation of \( \hat{F} \) under a \( \Lambda  \)-transformation is 
defined by
\begin{eqnarray}
\delta \hat{F}\left( \theta \right)  & = & \hat{F}\left( \theta 
	+\delta \theta \right) -\hat{F}\left( \theta \right) =g\delta \theta g
	-F\delta \theta F-A_{k}\delta \theta ^{kl}\partial _{l}F\nonumber \\
 & - & g\delta \theta gF-F\theta g\delta \theta g+\frac{1}{2}
	\left( g\delta \theta g\right) _{ks}\,x^{s}\theta ^{kl}
	\partial _{l}F+K_{\theta \delta \theta }\label{35} 
\end{eqnarray}
where the leading and subleading terms are the orders of \( \delta 
\theta ,\, \theta \delta \theta  \)
respectively, and \( K_{\theta \delta \theta } \) represents the 
\( \theta \delta \theta  \)
order terms coming from \( \mathcal{T}_{\left( \theta 
	+\delta \theta \right) ^{2}}-\mathcal{T}_{\theta ^{2}} \),
which can be obtained by replacing one of the \( \theta  \)'s 
by \( \delta \theta  \)
in \( \mathcal{T}_{\theta ^{2}} \). It can be written as 
\begin{eqnarray}
K_{\theta \delta \theta } & = & F\delta \theta F\theta F
	+F\theta F\delta \theta F+\frac{1}{2}A_{k}\delta \theta ^{kl}
	\left( \partial _{l}A_{m}+F_{lm}\right) 
	\theta ^{mn}\partial _{n}F\nonumber \\
 & + & \frac{1}{2}A_{k}\theta ^{kl}\left( \partial _{l}A_{m}+F_{lm}\right) 
	\delta \theta ^{mn}\partial _{n}F\nonumber \\
 & + & A_{k}\delta \theta ^{kl}\partial _{l}\left( F\delta F\right) 
   +A_{k}\theta ^{kl}\partial _{l}\left( F\delta \theta F\right) +\nonumber \\
 &  & \delta \theta ^{kl}\theta ^{mn}A_{k}A_{m}\partial _{l}
	\partial _{n}F.\label{36} 
\end{eqnarray}
Using eqs. (\ref{33})--(\ref{35}), \( \delta G+\delta \hat{F} \) 
can be expressed to subleading order as
\begin{equation}
\label{37}
\delta G+\delta \hat{F}=P_{1}\left( \delta \theta \right) 
	+P_{2}\left( \theta \delta \theta \right) ,
\end{equation}
where \( P_{1}\left( \delta \theta \right)  \) and 
	\( P_{2}\left( \theta \delta \theta \right)  \)
represent the terms with the orders of \( \delta \theta  \) 
	and \( \theta \delta \theta  \).
They are given by
\begin{equation}
\label{38}
P_{1}\left( \delta \theta \right) =g\delta \theta g-F\delta \theta F
	-A_{k}\delta \theta ^{kl}\partial _{l}F
\end{equation}
\begin{eqnarray}
P_{2}\left( \theta \delta \theta \right)  & = & - g\delta \theta g\theta 
   \left( g+F\right) -\left( g+F\right) \theta g\delta \theta g\nonumber \\
 & + & \frac{1}{2}\left( g\delta \theta g\right) _{ks}\,x^{s}
	\theta ^{kl}\partial _{l}F+F\delta \theta F\theta F
	+F\theta F\delta \theta F\nonumber \\
 & + & \frac{1}{2}A_{k}\delta \theta ^{kl}\left( \partial _{l}A_{m}
	+F_{lm}\right) \theta ^{mn}\partial _{n}F\nonumber \\
 & + & \frac{1}{2}A_{k}\theta ^{kl}\left( \partial _{l}A_{m}
	+F_{lm}\right) \delta \theta ^{mn}\partial _{n}F
	+A_{k}\delta \theta ^{kl}\partial _{l}\left( F\theta F\right) 
	\nonumber \\
 & + & A_{k}\theta ^{kl}\partial _{l}\left( F\delta \theta F\right) 
	+\delta \theta ^{kl}\theta ^{mn}A_{k}A_{m}\partial _{l}\partial _{n}F
\label{39} 
\end{eqnarray}
The matrix \( \left( G+\hat{F}\right) ^{-1} \) and the square root of the 
determinant
of \( G+\hat{F} \), written in terms of \( g \) and \( F \) 
to subleading order, are
\begin{eqnarray}
\frac{1}{G+\hat{F}} & = & \frac{1}{g+F}\left[ 1+\left( F\theta F
	+\theta ^{kl}A_{k}\partial _{l}F\right) \frac{1}{g+F}\right] 
	\nonumber \\
\det{}^{\frac{1}{2}}\left( G+\hat{F}\right)  & = & 
	\det{}^{\frac{1}{2}}\left( g+F\right) 
	\left\{ 1 - \frac{1}{2}\Tr\left[ \frac{1}{g+F}\left( F\theta F
	+A_{k}\theta ^{kl}\partial _{l}F\right) \right] \right\} \label{40} 
\end{eqnarray}
where the \( \theta ^{2} \) term in \( G \) is omitted.

The noncommutative DBI Lagrangian is given by \cite{sw1} 
\begin{equation}
\label{41}
\hat{\mathcal{L}}\, _{DBI}\left( \hat{A}\right) =\frac{1}{G_{s}}
	\det{}^{\frac{1}{2}}\left( G+\hat{F}\right) .
\end{equation}
The variation of \( \hat{\mathcal{L}}\, _{DBI} \) 
under a \( \Lambda  \)-transformation is
\begin{equation}
\label{42}
\delta \hat{\mathcal{L}}\, _{DBI}=\frac{1}{2G_{s}}\det{}^{\frac{1}{2}}
	\left( G+\hat{F}\right) \left[ -\frac{2\delta G_{s}}{G_{s}}
	+\Tr\frac{1}{G+\hat{F}}\left( \delta G+\delta \hat{F}\right) \right]. 
\end{equation}
Inserting (\ref{32}), (\ref{33}) and (\ref{37})--(\ref{40}) into
(\ref{42}), we have, to subleading order,
\begin{eqnarray}
\delta \hat{\mathcal{L}}\, _{DBI} & = & \frac{1}{2G_{s}}\det{}^{\frac{1}{2}}
	\left( g+F\right) \left\{ 1-\frac{1}{2}\Tr\left[ \left( F\theta F
	+\theta ^{kl}A_{k}\partial _{l}F\right) \frac{1}{g+F}\right] \right\} 
	\nonumber \\
 & & \times\Tr\left\{ \theta g\delta \theta g+\frac{1}{g+F}P_{1}
	\left( \delta \theta \right) +\frac{1}{g+F}\left[ 
	\strut	P_{2}
	\left( \theta \delta \theta \right)  \right. \right. \nonumber \\
 &  & \left. \left. ~~~~~~~~~~+\left( F\theta F
	+\theta ^{kl}A_{k}\partial _{l}F\right)
\frac{1}{g+F}P_{1}\left( \delta \theta \right) \right] 
	\right\} +\hbox{higher order terms}\nonumber \\
 & = & \frac{1}{2G_{s}}\left( \delta {\mathcal{L}}_{1}
	+\delta \mathcal{L}_{2}\right) +\hbox{higher order terms},\label{43} 
\end{eqnarray}
where \( \delta {\mathcal{L}}\, _{1} \) is of order \( \delta \theta  \),
and \( \delta {\mathcal{L}}\, _{2} \) 
is of order \( \theta \delta \theta  \). We find
\begin{equation}
\label{44}
\delta \mathcal{L}\, _{1}=\det{}^{\frac{1}{2}}\left( g+F\right) 
\Tr\frac{1}{g+F}P_{1}\left( \delta \theta \right),
\end{equation}
\begin{eqnarray}
\delta \mathcal{L}\, _{2} & = & \det{}^{\frac{1}{2}}\left( g+F\right) 
\left\{ -\frac{1}{2}\left[ \Tr\frac{1}{g+F}\left( F\theta F+\theta ^{kl}A_{k}
\partial _{l}F\right) \right] .\left[ \Tr\frac{1}{g+F}P_{1}\left( \delta 
\theta \right) \right] \right. \nonumber \\
 & + & \Tr\left. \left( \theta g\delta \theta g+\frac{1}{g+F}\left[ P_{2}
	\left( \theta \delta \theta \right) +\left( F\theta F
	+\theta ^{kl}A_{k}\partial _{l}F\right) \frac{1}{g+F}P_{1}
	\left( \delta \theta \right) \right] \right) \right\} \label{45} .
\end{eqnarray}

First we examine the leading order term \( \delta {\mathcal{L}}\, _{1} \).
Inserting (\ref{38}) into (\ref{44}) we get
\begin{eqnarray}
\delta \mathcal{L}\, _{1} & = & \det{}^{\frac{1}{2}}\left( g+F\right) 
\Tr\frac{1}{g+F}\left( g\delta \theta g-F\delta \theta F-A_{k}\delta 
\theta ^{kl}\partial _{l}F\right) \nonumber \\
 & = & \det{}^{\frac{1}{2}}\left( g+F\right) \Tr\left[ \frac{1}{g+F}
	\left( g\delta \theta g-F\delta \theta F\right) 
	+\delta \theta F\right] \nonumber \\
 & + & \hbox{total derivative},\label{46} 
\end{eqnarray}
where in the second step we used
\begin{equation}
\label{47}
	\det{}^{\frac{1}{2}}\left( g+F\right) \Tr\frac{1}{g+F}\partial _{l}F
	=2\partial _{l}\det{}^{\frac{1}{2}}\left( g+F\right) 
\end{equation}
and then integrated by parts. Eq. (\ref{46}) can be further reduced to
\begin{eqnarray}
\delta \mathcal{L}\, _{1} & = & \det{}^{\frac{1}{2}}
	\left( g+F\right) \Tr\left( \delta \theta g\right) 
	+\hbox{total derivative}\nonumber \\
 & = & \hbox{total derivative},\label{48} 
\end{eqnarray}
where we have exploited that \( \delta \theta  \) is an antisymmetric matrix 
and that \( g \)
is symmetric, thus \( \Tr\left( \delta \theta g\right) = 0 \).

Eq. (\ref{48}) shows that under a \( \Lambda  \)-transformation the leading 
order term of the variation of the noncommutative DBI Lagrangian is a
total derivative.

Next we consider the subleading order term 
\( \delta {\mathcal{L}}\,_{2} \). Plugging (\ref{38}) and (\ref{39}) 
into (\ref{45}) we get					\def\igpF{{1\over g+F}}
\begin{eqnarray}
\! \delta\mathcal{L}_{2}\!&\!=\!&\det{}^\2(g+F)\Biggl\lbrace-\2\left[\Tr\igpF
	(g\d\th g-F\d\th F-A_k\d\th^ {kl}\6_lF)\right]
 	\left[\Tr (F\th F+\th^{kl}A_k\6_lF)\igpF\right]	\HS-19
\nn\\&&	+ \Tr\Bigl(\th g\d\th g+\igpF\Bigl[ -g\d\th g\th(g+F)-(g+F)\th g\d\th g
	+\2(g\d\th g)_{ks}\,x^s\th^{kl}\6_lF 
\nn\\&&	+ F\d\th F\th F+ F\th F\d\th F 
	+\2A_k\d\th^{kl}(\6_lA_m+F_{lm})\th^{mn}\6_nF
	 +\2A_k\th^{kl}(\6_lA_m+F_{lm})\d\th^{mn}\6_nF	\HS-19
\nn\\&&	+A_k\d\th^{kl}\6_l(F\th F)+A_k\th^{kl}\6_l(F\d\th F)
	 +\d\th^{kl}\th^{mn}A_kA_m\6_l\6_nF\Bigr]
\nn\\&&	+\igpF(F\th F+\th^{kl}A_k\6_lF)\igpF(g\d\th g
	-F\d\th F-A_m\d\th^{mn}\6_nF)\Bigr)\Biggr\rbrace	\label{49}
\end{eqnarray}
\( \delta \mathcal{L}\, _{2} \) seems quite complicated. The strategy that
we will use to simplify \( \delta {\mathcal{L}}\, _{2} \) is to 
exploit (\ref{47}) and the equation
\begin{eqnarray}
2\partial _{l}\partial _{n}\det{}^{\frac{1}{2}}\left( g+F\right)  
	& = & \det{}^{\frac{1}{2}}\left( g+F\right) \left[ \frac{1}{2}
	\left( \Tr\frac{1}{g+F}\partial _{l}F\right) 
	\left( \Tr\frac{1}{g+F}\partial _{n}F\right) \right. \nonumber \\
 & - & \left. \Tr\left( \frac{1}{g+F}\partial _{l}F\frac{1}{g+F}\partial _{n}F
	\right) +\Tr\frac{1}{g+F}\partial _{l}\partial_{n}F\right], \label{50}
\end{eqnarray}
and then integrate by parts, keeping terms like 
\( \partial _{l}\left( F\theta F\right)  \) unchanged. Then we have
\begin{eqnarray}
\delta \mathcal{L}\, _{2} & = & \det{}^{\frac{1}{2}}\left( g+F\right) 
\left\{ -\frac{1}{2}\left[ \Tr\frac{1}{g+F}\left( g\delta \theta g-F\delta 
\theta F\right) \right] .\left( \Tr\frac{1}{g+F}F\theta F\right) \right. 
\nonumber \\
 & + & \partial _{l}\left[ A_{k}\theta ^{kl}\Tr\frac{1}{g+F}
	\left( g\delta \theta g-F\delta \theta F\right) \right] -\partial _{l}
	\left( A_{k}\delta \theta ^{kl}\Tr\frac{1}{g+F}F\theta F\right) 
	\nonumber \\
 & + & 2\partial _{l}\partial _{n}\left( A_{k}A_{m}\delta \theta ^{kl}
	\theta ^{mn}\right) -\Tr\left( \theta g\delta \theta g\right) 
	\nonumber \\
 & - & \left( g\delta \theta g\right) _{ks}\left( \partial _{l}
	\,x^{s}\right) \theta ^{kl}+\Tr\frac{1}{g+F}
	\left[ F\delta \theta F\theta F+F\theta F\delta \theta F\right. 
	\nonumber \\
 &  & ~~~~~~~~~~~+\left. A_{k}\delta \theta ^{kl}\partial _{l}
		\left( F\theta F\right) +A_{k}\theta ^{kl}\partial _{l}
		\left( F\delta \theta F\right) \right] \nonumber \\
 & - & \delta \theta ^{kl}\theta ^{mn}\partial _{n}\left[ A_{k}
	\left( 2\partial _{l}A_{m}-\partial _{m}A_{l}\right) \right] 
  -  	\theta ^{kl}\delta \theta ^{mn}\partial _n\left[ A_{k}
	\left( 2\partial _{l}A_{m}-\partial _{m}A_{l}\right) \right] 
	\nonumber \\
 & + & \Tr\frac{1}{g+F}F\theta F\frac{1}{g+F}\left( g\delta \theta g
	-F\delta \theta F\right) 
   -   \Tr\frac{1}{g+F}F\theta F\frac{1}{g+F}A_{k}\delta \theta ^{kl}
	\partial _{l}F\nonumber \label{51} \\
 & + & \left. \Tr\frac{1}{g+F}\theta ^{kl}A_{k}\partial _{l}F\frac{1}{g+F}
	\left( g\delta \theta g-F\delta \theta F\right) \right\} 
	+\hbox{total derivative}.
\end{eqnarray}
After a straightforward calculation with many cancellations 
we are left with
\begin{eqnarray}
\delta \mathcal{L}\, _{2} & = & \Tr\left[ -F\delta \theta F\theta 
	+\frac{1}{g+F}\left( F\delta \theta F\theta F
	+F\theta F\delta \theta F\right) \right. \nonumber \\
 & + & \left. \frac{1}{g+F}F\theta F\frac{1}{g+F}
	\left( g\delta \theta g-F\delta \theta F\right) \right] 
	+\hbox{total derivative}.\label{52} 
\end{eqnarray}
If we rewrite \( g\delta \theta g-F\delta \theta F \) as
\begin{equation}
\label{53}
g\delta \theta g-F\delta \theta F=-\left( g+F\right) \delta \theta 
	\left( g+F\right) +\left( g+F\right) \delta \theta g+g\delta \theta 
	\left( g+F\right) 
\end{equation}
we can further simplify (\ref{52}) and obtain
\begin{equation}
\label{54}
\delta \mathcal{L}\, _{2}=\hbox{total derivative},
\end{equation}
which shows that the subleading order term of the variation of the 
	noncommutative
DBI Lagrangian is a total derivative under \( \Lambda  \)-transformations.

{}From the above calculation we have seen that the check of the 
\( \Lambda  \)-symmetry
for the noncommutative DBI action is high nontrivial, and we believe
that if the ordinary gauge theory has \( \Lambda  \)-symmetry, the 
corresponding
noncommutative gauge theory should also possess this symmetry.

\section{Summary and discussions}

So far we have examined the \( \Lambda  \)-symmetry of noncommutative 
Yang-Mills
theory on \(\mathbb R^{n} \). We have calculated the variation of the 
noncommutative
Yang-Mills Lagrangian including the subleading order terms. We have found 
that the leading
and subleading order terms in the variation of the Lagrangian are total 
derivatives,
which indicates that the noncommutative Yang-Mills action is background 
independent
if the ordinary gauge field damps fast enough at infinity. In the context
of  noncommutative Yang Mills theory on ordinary spacetime, we have shown
that the quantity \( Q \) defined in \cite{sw1} is not background independent
for a general gauge field \( A \). Only if the field strength \( F \) takes
the constant value, the quantity \( Q \) is background independent. For the
noncommutative DBI theory we have found that for small \( B \)
the variation of the noncommutative DBI
Lagrangian under a \( \Lambda  \)-transformation
is a total derivative  to subleading order. Thus, when the ordinary
gauge field damps fast enough at the infinity, the noncommutative DBI
action possesses  \( \Lambda  \)-symmetry, which is different from 
the ordinary
DBI theory, where the Lagrangian is manifestly \( \Lambda  \)-symmetric.
Since our check for  \( \Lambda  \)-symmetry of  noncommutative gauge
theories, which act as the low energy effective theories of \( D \)-branes,
is highly non-trivial, we conclude that if the ordinary gauge theory has
\( \Lambda  \)-symmetry, its noncommutative counterpart should also possess
this symmetry.

In the present paper, the \( \Lambda  \)-transformation is realized in the 
framework of
 ordinary gauge fields, that is, we express \( \delta \hat{F} \) in terms of
\( A,\, F \) and \( \theta  \), so that our discussion has to be restricted to
small values of \( \theta  \). It would be interesting to see whether it is
possible to discuss \( \Lambda  \)-transformations on noncommutative gauge 
field
directly. In \cite{s6} it has been shown that \( \Lambda  \)-symmetry could
be modified in the presence of the \( B- \)field, similarly to the 
\( \lambda  \)-symmetry.
Under this deformed \( \Lambda  \)-transformation, the \( B \)-field 
transforms as: 
\( B\rightarrow B + d\Lambda +i\{\Lambda,A\}_{MB} . \)
It would be interesting
to study the relation between the deformed \( \Lambda  \)-transformation and
the present one explicitely.

Recently, a hybrid point splitting regularization that leads to 
the Seiberg-Witten
description including the general two-form \( \Phi  \) was found \cite{ad7}.
This suggests an investigation about how the \( \Lambda  \)-transformation 
works in
the presence of the general two-form \( \Phi  \), since in this case we have
more freedom for the  description \cite{sw1}, which now can be realized 
within
the standard renormalization scheme by some freedom in changing variables. We
hope to answer these questions in the near future.

\bigskip{}
\noindent \textbf{\Large Acknowledgements}{\Large \par}

\bigskip{}
\noindent We would like to thank K. Okuyama for valuable
discussion. This work is supported in part by the {\it Austrian 
Research Funds} FWF under grant Nr. M535-TPH.

\end{document}